\documentclass[preprint,12pt]{elsarticle}
\usepackage{amssymb}
\usepackage{caption} 
\journal{xxx}

\begin{document}

\begin{frontmatter}


\title{First-Principles Investigation of Grain Boundary Effects on Fluorine-Induced Initial Corrosion of NiCr Alloys}

\author[1]{Hamdy Arkoub}
\author[1]{Miaomiao Jin\corref{cor}}
\affiliation[1]{Department of Nuclear Engineering, The Pennsylvania State University, 218 Hallowell Building, University Park, PA 16802, USA}
\cortext[cor]{Corresponding author: mmjin@psu.edu}

\begin{abstract}
Chromium depletion at grain boundaries (GBs) due to selective attack is a critical issue in the molten salt corrosion of NiCr alloys. Despite the importance of GBs in this process from numerous experimental studies, most theoretical work has predominantly focused on fluorine interactions with idealized crystalline surfaces, neglecting the complexity of GB local environments. This study aims to bridge that gap by employing density functional theory (DFT) to investigate the atomic interactions and Cr dissolution mechanisms at GB in NiCr alloys under molten fluoride salt environments. Specifically, a $\Sigma$5(210)/(001) symmetrical tilt GB is constructed to explore the adsorption energies of fluorine on Ni(100) and Cr-doped Ni(100) surfaces. We find that fluorine exhibits a strong preference for binding at GB sites, with Cr doping amplifying this effect, leading to higher adsorption energies compared to bulk Ni surfaces. Fluorine bonding with Cr significantly alters the interaction between Cr-F complexes and Ni substrate, and the consequent dissolution barriers for Cr atoms; the formation of CrF$_3$ largely reduces the energy barrier for Cr dissolution. This work highlights the essential role of GBs in enhancing fluorine adsorption and accelerating Cr depletion, providing new insights into the mechanisms of early-stage corrosion in NiCr alloys.
\end{abstract}


\end{frontmatter}

\section{Introduction}
\label{sec:sample1}

The renewed interest in utilizing molten fluoride salts as fuel and coolant in molten salt reactors (MSRs), given their favorable chemical stability and thermophysical characteristics at high-temperature \cite{williams2006assessment,delpech2010molten}, necessitates a comprehensive understanding of the intricate metal-salt interactions occurring at the interface \cite{leblanc2010molten,kelly2014generation,startt2019modeling}. The high-temperature molten fluoride salts possess high corrosivity since the passive oxide films have been found to be chemically unstable under the molten salt environment \cite{sohal2010engineering}. Nickel-based alloys, due to their good high-temperature mechanical properties, sufficient resistance to neutron radiation, and high corrosion resistance, are accepted as the candidate structural materials in MSR \cite{zohuri2020generation}. However, the high-temperature corrosion observed in molten-salt systems is most often facilitated by the depletion of active elements in the alloy along surfaces and grain boundaries (GBs), and with the degree of corrosion closely tied to the Cr concentration in the alloys \cite{olson2009materials,pint2019re,leong2023kinetics, arkoub2024reactive}. The corrosion process involves intricate chemical reactions occurring at the alloy-salt interface, where corrosive species (fluorine, F) in molten salts react with active alloying elements, particularly Cr, that diffuse from the bulk to the surface of the alloys \cite{williams2006assessment,olson2009materials,arkoub2024reactive}, leading to Cr dissolution.

Surface adsorption and diffusion of fluorine play a crucial role in the early stages of corrosion in NiCr alloys. First-principles studies have provided insights into these mechanisms, particularly in terms of Cr segregation and dissolution. For instance, Ren et al. \cite{ren2016adsorption} demonstrated that Cr acts as a trap site for fluorine on Ni(111) surface, facilitating its diffusion and enhancing Cr dissolution. Similarly, Yin et al. \cite{yin2018first, yin2018theoretical} highlighted how increased fluorine coverage alters surface morphology, promoting selective Cr dissolution in the form of CrF$\mathrm{_{2}}$/CrF$\mathrm{_{3}}$ molecules in the salt. These findings emphasize the strong interaction between fluorine and chromium, driving the initial stages of corrosion. Additionally, Startt et al. \cite{startt2021ab} showed that while Cr tends to remain within the bulk under vacuum conditions, the presence of fluorine significantly enhances Cr segregation to the surface due to valence charge transfer. Recent experimental and phase field model predictions by Mills et al. \cite{mills2024elucidating} confirmed that Cr enriches the salt near the interface when NiCr is exposed to molten FLiNaK, reflecting the thermodynamic preference for Cr dissolution over Ni.

An intriguing aspect of molten fluoride salt corrosion in NiCr alloys is the tendency for corrosion to predominantly occur at GBs, where Cr dissolution is diffusion-limited \cite{olson2009materials,sridharan2013corrosion,guo2018corrosion,zheng2018corrosion}. GBs serve as preferential pathways for diffusion, making them particularly vulnerable to corrosion initiation. Th GB-facilitated atom diffusion is strongly influenced by grain size and orientation, with finer grains accelerating Cr dissolution due to the effective faster diffusion of Cr along GBs compared to bulk lattice diffusion \cite{wang2016effect,zheng2015corrosion}. High-angle GBs, in particular, play a critical role in molten salt corrosion \cite{maric2018effect}. The more disordered atomic structures and higher free energy compared to the bulk lattice \cite{rohrer2011grain} render atoms at GBs more susceptible to dissolution \cite{bawane2022visualizing,badwe2018decoupling}, thereby accelerating intergranular corrosion. This GB-driven dissolution can result in pitting and void formation along GBs, where salt can accumulate \cite{mills2024elucidating,yin2024europium,gill2021investigating,ai2018influence,yin2018effect}. Ultimately, intergranular corrosion may lead to the formation of continuous wormhole pores \cite{yang2023one}, while voids from bulk dealloying can create discontinuous pores \cite{liu2021formation,liu2023temperature}. Recent work by Zhou et al. \cite{zhou2020proton} investigated the corrosive behavior of NiCr alloys in contact with FLiNaK molten salt and demonstrated a strong preference for salt to attack GBs, selectively leaching Cr from these regions. The excess energy at GBs likely accelerates dealloying via a GB wetting mechanism in the liquid phase, as proposed by Joo et al. \cite{joo2022inhomogeneous}. Interestingly, Zhou et al. also found that proton irradiation decelerates intergranular corrosion in NiCr alloys; they attributed it to irradiation-enhanced bulk diffusion, which increases the migration of Ni and Cr atoms from the grain interior to the GBs to mitigate void formation at the boundaries, thereby slowing down the corrosion process. This is in contrast with another work by Bakai et al. \cite{bakai2008combined}, where they observed an increased intergranular corrosion rate in Ni-Mo alloys exposed to NaF-ZrF$_4$ salt under electron irradiation; it was suggested that radiation-enhanced diffusion of Cr toward the interface, depletes protective elements within the bulk and increases susceptibility to corrosion. These conflicting findings underscore the crucial influence of atomic kinetics and GB chemistry on the corrosion response.

Despite the importance of GBs in molten fluoride salt corrosion from numerous experimental studies, theoretical work typically focused on fluorine interactions with idealized surfaces to understand the atomic-level corrosion mechanisms. Hence, the localized interactions and bonding nature of fluorine with NiCr alloy at GBs remain to be elucidated. In this study, we aim to address this gap by considering the interaction behaviors of fluorine (F) with metal surfaces at the GB region in NiCr alloys. To accomplish this, we employ Density Functional Theory (DFT) to probe electronic structure and atomic interactions with a $\mathrm{\Sigma}$5(210)/(001) symmetrical tilt GB model. We will first demonstrate the adsorption energetics of F on the Ni and Cr-doped Ni GBs to assess the impact of GBs on fluorine adsorption. Next, we will clarify the role of GBs in Cr dissolution energetics, particularly in the presence of fluorine. Finally, we will analyze Cr dissolution under different conditions to better understand GB-driven corrosion mechanisms in NiCr alloys.  

\section{Computational Details}

First-principles calculations are performed using DFT as implemented in the Vienna Ab initio Simulation Package (VASP) \cite{kresse1993ab,kresse1996efficient}. Electron-ion interactions are described using projector-augmented wave (PAW) potentials, and exchange-correlation effects are treated with the generalized gradient approximation (GGA) in the Perdew-Burke-Ernzerhof (PBE) form \cite{perdew1996k,blochl1994projector,kresse1999ultrasoft}. Geometry optimization is conducted using the conjugate-gradient scheme, with total energy convergence set to 10$^{-6}$ eV. The valence electrons considered are Ni 4s$^2$3d$^8$, Cr 4s$^1$3d$^5$, and F 2s$^2$2p$^5$, and spin polarization effects are included in all calculations. A plane wave energy cutoff of 400 eV is applied.

\begin{figure}[!ht]
	\centering
	\includegraphics[width=0.6\textwidth]{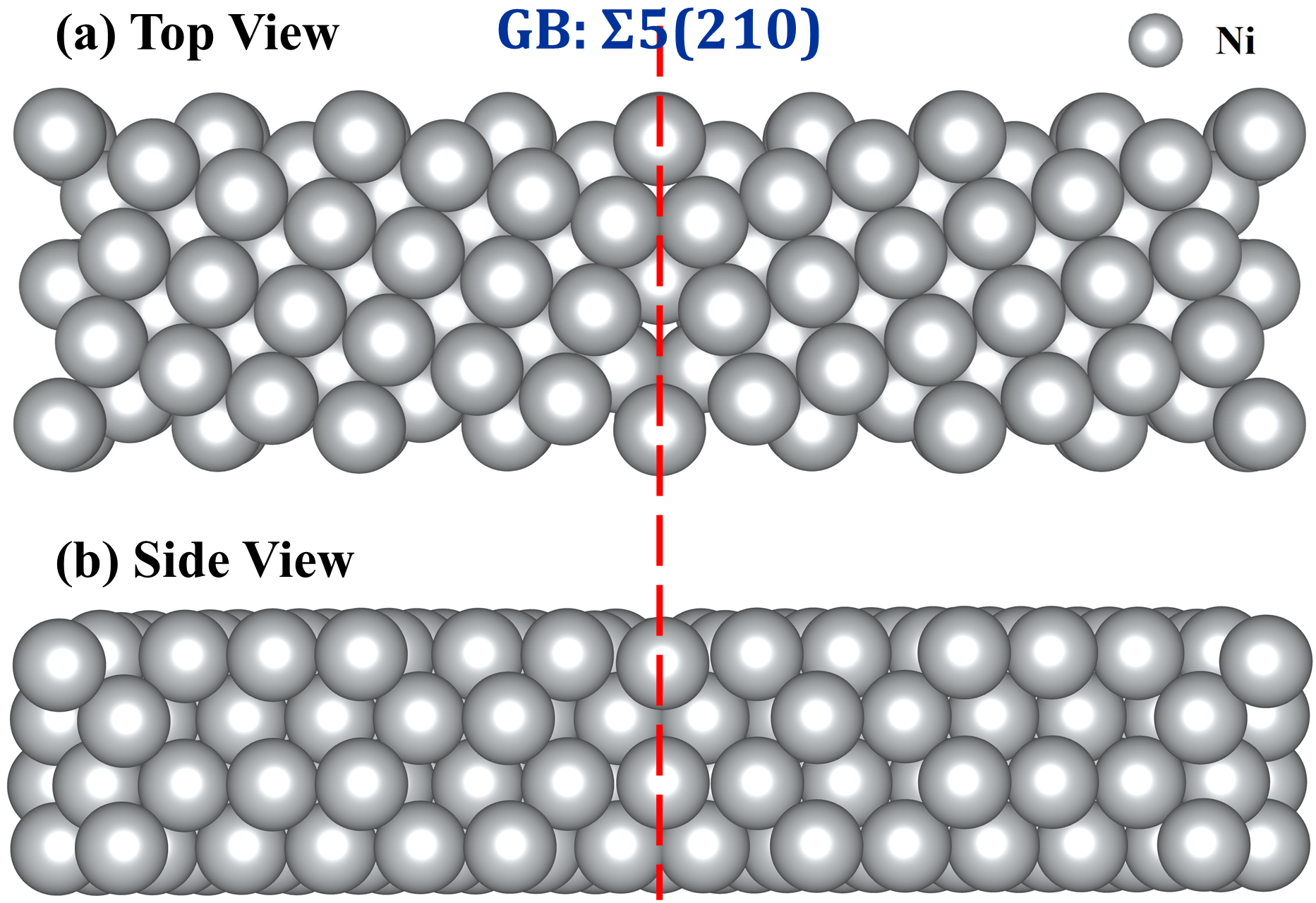}
	\caption{(a) Top and (b) side view of the optimized symmetric tilt GB used in the study.}
	\label{fig:model}
\end{figure}

For this study, a $\mathrm{\Sigma}$5(210)/(001) symmetrical tilt GB was constructed. This GB model with a rotational angle of 36.9$^\mathrm{o}$ has an energy of approximately 1265 mJ/m$^2$ for the $\langle100\rangle$ tilt GBs in polycrystalline Ni \cite{sangid2010grain}. The optimized structure of the STGB is shown in Figure \ref{fig:model}. The supercell system is periodic in all directions, and a slab structure containing four Ni layers (40 Ni atoms in each layer) is used to simulate the (100) surface. To reduce interactions between repeated slabs, a vacuum layer of 24.41 {\AA} was added. The bottom layer of the slab was kept frozen while the remaining layers were allowed to relax. The relaxed supercell dimensions are 32.3 \AA $\times$ 7.8 \AA $\times$ 30.0 \AA. A Gamma-centered scheme were used for Brillouin zone integration, with a k-point mesh of 1 $\times$ 3 $\times$ 1. Convergence tests were conducted to ensure that the spacing between the GBs is large enough to minimize the interaction between adjacent boundaries and represent bulk-like conditions in the center of the grain. Additionally, the width of the GB model was confirmed to provide accurate results (details in the Supplementary Materials (SM)). To model a Cr-doped Ni surface, one Ni atom at the surface was substituted by a Cr atom, a well-adopted method for studying the catalytic properties of binary alloys \cite{kyriakou2012isolated,tierney2009hydrogen}. Atom visualization was accomplished using the VESTA program \cite{momma2011vesta}.

The F adsorption energy ($E_\mathrm{{ads}}$) at the different adsorption sites is calculated using the following equation:

\[E_{\mathrm{ads}}=E_{\mathrm{F/sub}}-E_{\mathrm{sub}}-0.5 \times E_{\mathrm{F_{2}}}\qquad(1)\] 
where $E_{\mathrm{F_{2}}}$, $E_{\mathrm{sub}}$, $E_{\mathrm{F/sub}}$ are the energies of the F$\mathrm{_2}$ molecule in vacuum, metal slab, and the system with fluorine adsorbed on the slab, respectively. With this definition, a more negative value of $\mathrm{E_{ads}}$ indicates stronger fluorine binding to the surface. To further analyze the thermodynamic favorability and stability of fluorine chemisorption, the average chemisorption energy per fluorine atom ($\mathrm{E_{chem}}$) is calculated as:

\[E_{\mathrm{chem}}=\frac{1}{\mathrm{N_{F}}}[E_{\mathrm{F/slab}}-(E_{\mathrm{slab}}+\mathrm{\frac{N_{F}}{2}}E_{\mathrm{F_{2}}})]\qquad(2)\]
where $E_{\mathrm{F/slab}}$ and $E_{\mathrm{slab}}$ are the total energies of the adsorbate–substrate system and the clean NiCr slab, respectively, and $N_{\mathrm{F}}$ is the number of adsorbed fluorine atoms. Like the adsorption energy, a more negative value of $E_{\mathrm{chem}}$ implies more favorable fluorine adsorption.

For the dissolution energy barrier ($E_\mathrm{{diss}}$), the calculation involves incrementally raising the Cr atom by 0.5 {\AA} above the surface and relaxing the configuration along the lateral directions (perpendicular to the surface normal). The dissolution energy is evaluated using the formula: 

\[E_{\mathrm{diss}}=E_{\mathrm{trans/sub}}-E_{\mathrm{initial/sub}}\qquad\qquad(3)\]
where $E_{\mathrm{initial/sub}}$ and $E_{\mathrm{trans/sub}}$ are the total energies for the initial state and transition state of Cr with respect to the metal substrate, respectively. The energy profile was traced to obtain these quantities.

\section{Results and Discussion}

\subsection{F Adsorption}

\begin{figure}[!ht]
	\centering
	\includegraphics[width=0.7\textwidth]{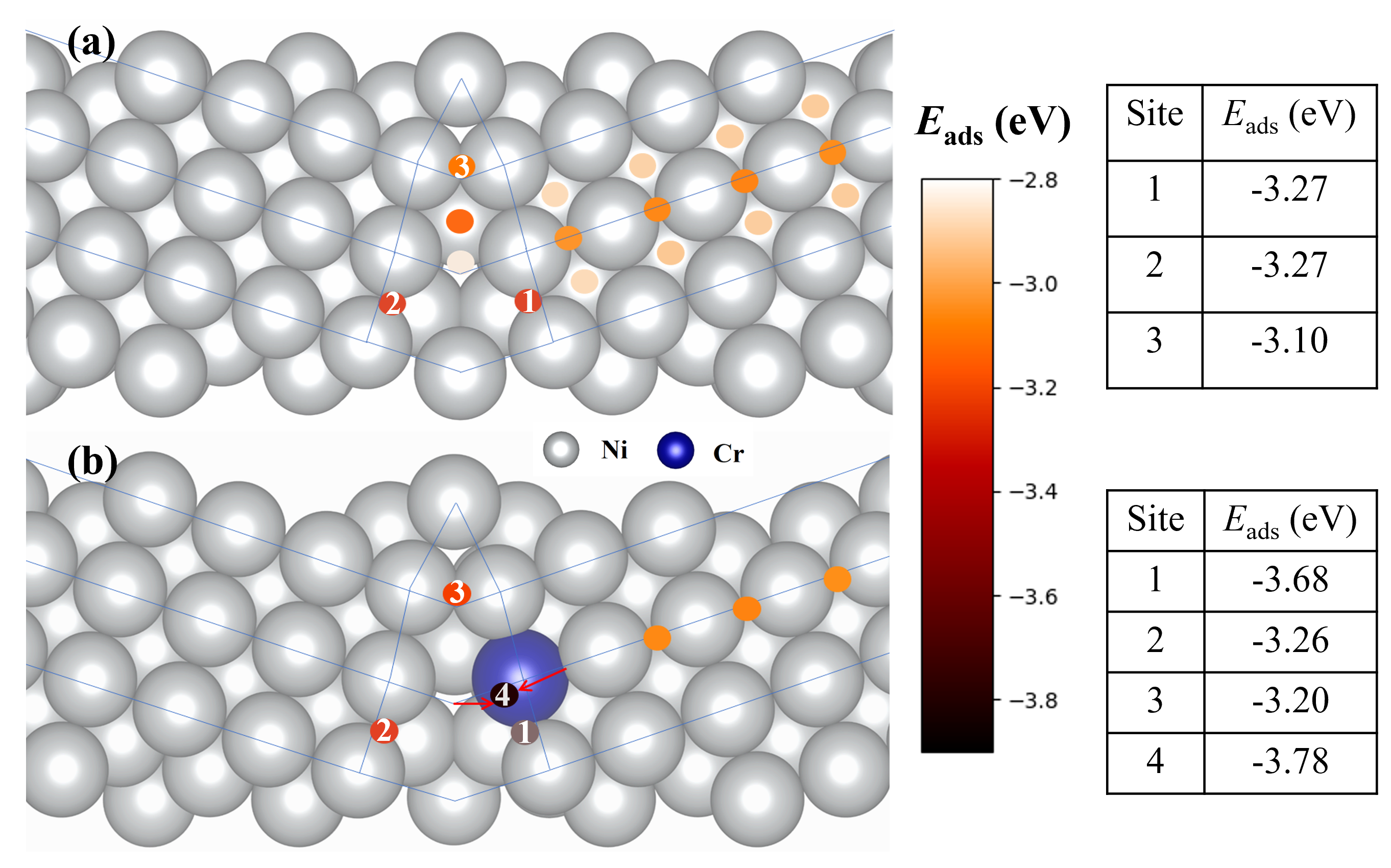}
	\caption{The fluorine adsorption sites and energies calculated on top of (a) Ni and (b) Cr-doped Ni GB surfaces.}
	\label{fig:ads_E}
\end{figure}

The initial step involves studying the adsorption behavior of fluorine on the GB surface. Given the presence of multiple surface sites at the GB, all potential F adsorption sites have been considered to identify any trends influenced by the GB's presence. The adsorption energies ($E_\mathrm{{ads}}$) of fluorine on both Ni GB and Cr-doped Ni GB surfaces are calculated using Eq. (1) to determine the energetically favorable sites for fluorine binding. A detailed breakdown of the adsorption sites and their respective energies for both Ni GB and Cr-doped Ni GB surfaces is provided in Figure \ref{fig:ads_E}. 

First, the F adsorption energies on pure Ni surface (Figure \ref{fig:ads_E}(a)) away from the GB, show strong agreement with the values reported by Yin et al.\cite{yin2018theoretical} for a single crystal Ni(100) surface. Specifically, the bridge site, identified as the most stable adsorption site, and the hollow site exhibit adsorption energies of 3.090 eV and 2.968 eV, respectively, which closely match the literature values of 3.09 eV and 2.97 eV \cite{yin2018theoretical}. However, moving towards the GB, the absolute adsorption energy increase further, ranging from 3.145 eV to 3.276 eV, indicating stronger fluorine binding along the GB. Notably, sites 1 and 2 emerge as the most stable adsorption sites within this region. As seen in Figure \ref{fig:ads_E}(a), these sites are shifted toward the GB center line rather than being located exactly at the expected bridge positions. Interestingly, the fluorine does not favor the interstitial sites directly at the GB. Overall, these adsorption values suggest that fluorine atoms exhibit a preference for adsorption at particular sites in the GB region, where the local atomic arrangement at the GB make these sites more energetically favorable compared to the bulk surface.

For the Cr-doped Ni surface, as shown in Figure  \ref{fig:ads_E}(b), the fluorine adsorption energies away from the GB are slightly higher than those on the pure Ni surface, with an average bridge site absolute adsorption energy of 3.096 eV. However, along the GB, the absolute adsorption energies at sites 2 and 3 increase further, reaching 3.200 eV and 3.259 eV, respectively, compared to the non-doped surface. This suggests that the Cr-doped Ni surface enhances fluorine adsorption compared to pure Ni, a trend consistent with previous findings in the literature based on (100) surfaces \cite{ren2016adsorption, yin2018theoretical}. 
Notably, near the Cr atom (sites 1 and 4), site 4 emerges as the most stable for fluorine adsorption, with fluorine atoms placed at nearby sites tending to slide toward this location. Additionally, a metastable site is identified at site 1, with adsorption energies of -3.68 eV and -3.78 eV for sites 1 and 4, respectively. These values are close to previous calculations for a perfect Cr-doped Ni surface, where fluorine adsorption at slightly off-top Cr sites resulted in adsorption energies around -3.75 eV \cite{ren2016adsorption, yin2018theoretical}. It suggests that the presence of Cr exerts a stronger influence than the GB itself in attracting fluorine to the surface, further enhancing the adsorption strength in the vicinity of Cr-doped GB regions.

\subsection{Cr Dissolution}

\subsubsection{F-free Cr Dissolution}

\begin{figure}[!ht]
	\centering
	\includegraphics[width=1.0\textwidth]{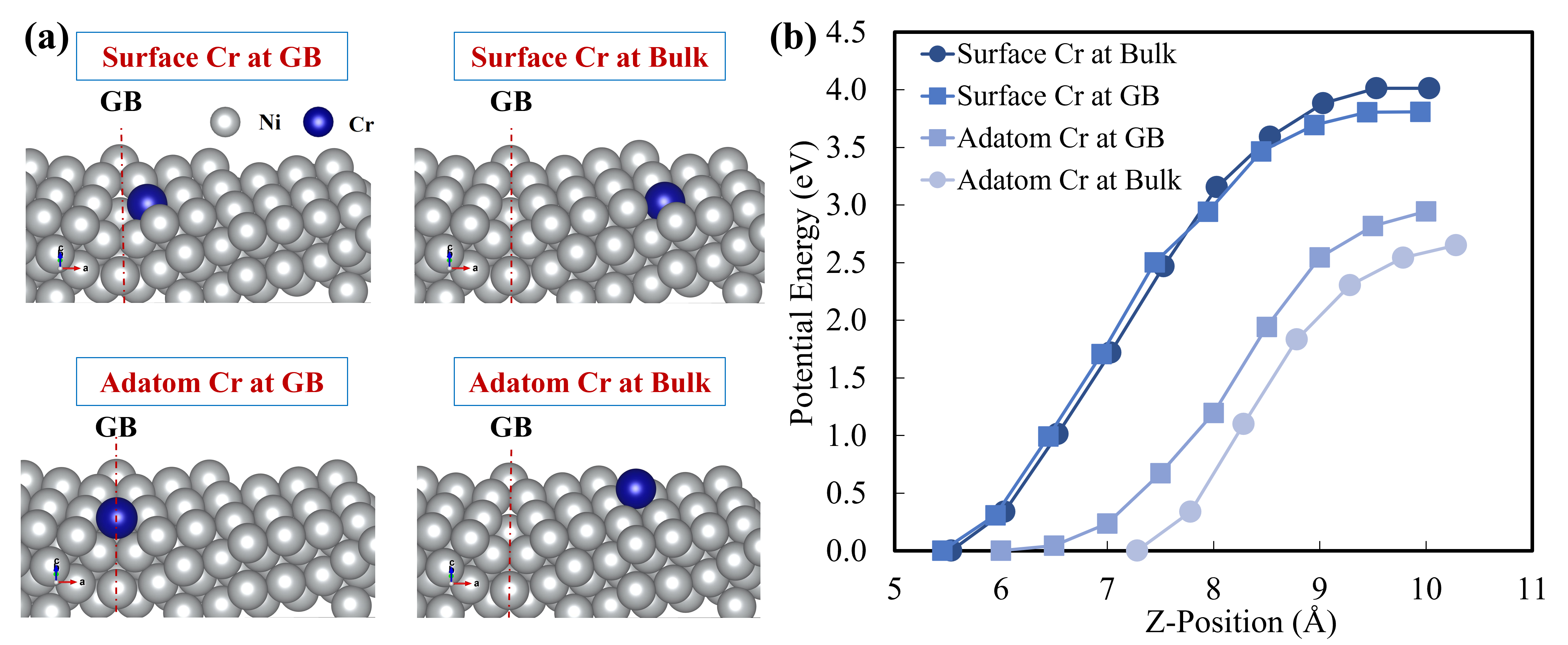}
	\caption{(a) Initial structures of different configurations of GB surface slab after optimization used for calculations of Cr dissolution barrier at and away from the GB. (b) The corresponding potential energy profiles for the dissolution of a Cr atom in vacuum.}
	\label{fig:diss_1}
\end{figure}

To simulate the dissolution of a Cr ion from both the GB and bulk surfaces in a vacuum, four scenarios were considered, as illustrated in Figure \ref{fig:diss_1}(a). Two cases involve the dissolution of a surface Cr atom: one from the bulk region (away from the GB) and the other at the GB. The other two cases involve a Cr atom adsorbed at the GB (Cr adatom at GB) and a Cr adatom in the bulk part (away from the GB). Note that in the case of the Cr adatom at the GB, the Cr sinks slightly into the GB due to the local interactions with the GB structure. In each scenario, the Cr atom is incrementally displaced from the surface in the absence of any solvent representation (i.e., in a vacuum). The dissolution energy profiles for all four cases are presented in Figure \ref{fig:diss_1}(b), allowing for a comparison of the dissolution behavior at both GB and bulk regions.

As the Cr atom dissociates from the metal slab, the system's energy steadily increases upon bond breaking, eventually plateauing at around 9 {\AA} in distance from the bottom of the supercell (Figure \ref{fig:diss_1}b). The energy profiles and dissolution energy barriers for surface Cr atoms in both cases, i.e., at the GB and in the bulk, are quite similar, with the GB dissolution barrier being slightly lower than that of the bulk (3.806 eV versus 4.013 eV). This indicates that while the GB provides a marginally easier path for Cr dissolution, the overall difference is not substantial for surface Cr atoms. However, a key observation emerges when considering the Cr adatom, which exhibits a significantly reduced dissolution barrier. The Cr adatom in the bulk shows the lowest barrier (2.650 eV), while the GB adatom has a slightly higher barrier (2.945 eV). This minor difference is attributed to the fact that the GB adatom is slightly absorbed into the open structure of the GB. Given that high-angle GBs are generally perceived as high-diffusion pathways for mass transport \cite{balluffi1982grain,atkinson1984diffusion}, this comparison suggests that GBs can serve as highly favorable avenue for Cr dissolution. Additionally, Cr atoms diffusing from the GB region can migrate to the bulk, where they can dissolve more easily as bulk adatoms. Such enhancement of localized corrosion underscores the critical role of GB in promoting Cr depletion during the early stages of corrosion in NiCr alloys. However, these calculations do not yet account for the influence of fluorine, which will be explored in the next section to understand its impact on Cr dissolution behavior.

\subsubsection{F-assisted Cr Dissolution}

\begin{figure}[!ht]
	\centering
	\includegraphics[width=1.0\textwidth]{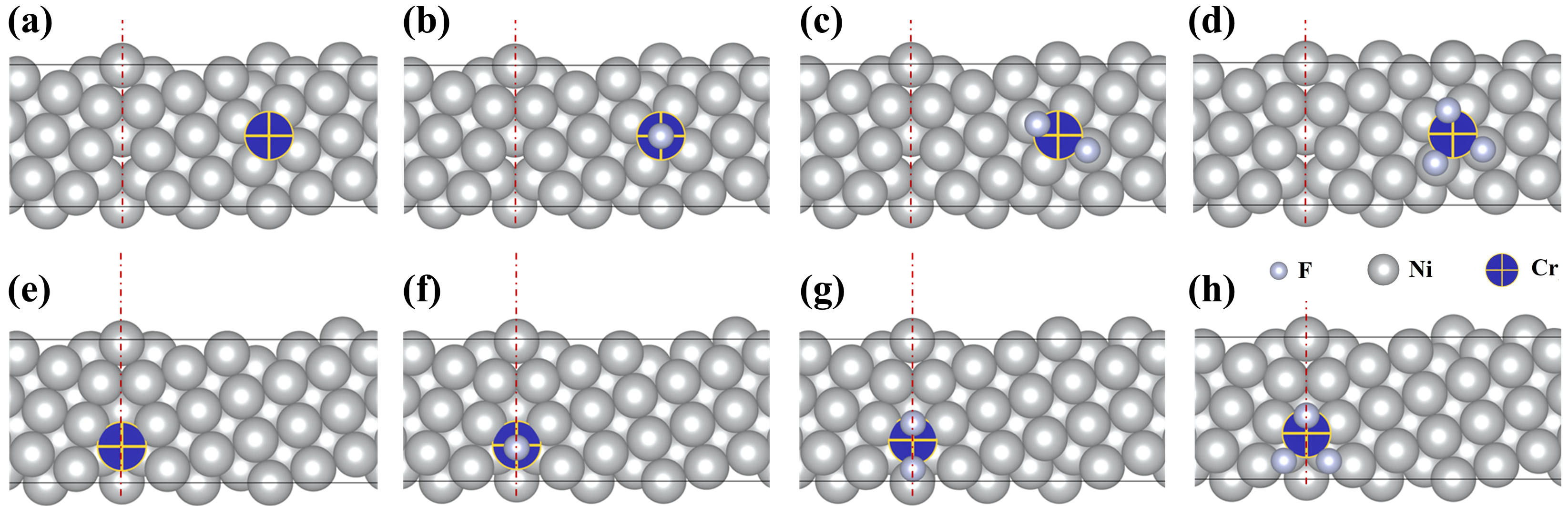}
	\caption{Initial configurations of the slab used for calculations of the Cr adatom dissolution barrier. The dashed red line indicates the GB. (a)–(d) represent configurations of Cr, CrF, CrF$_2$, and CrF$_3$ at the bulk, while (e)–(h) represent configurations of Cr, CrF, CrF$_2$, and CrF$_3$ at the GB.}
	\label{fig:adatom}
\end{figure}

\begin{figure}[!ht]
	\centering
	\includegraphics[width=0.6\textwidth]{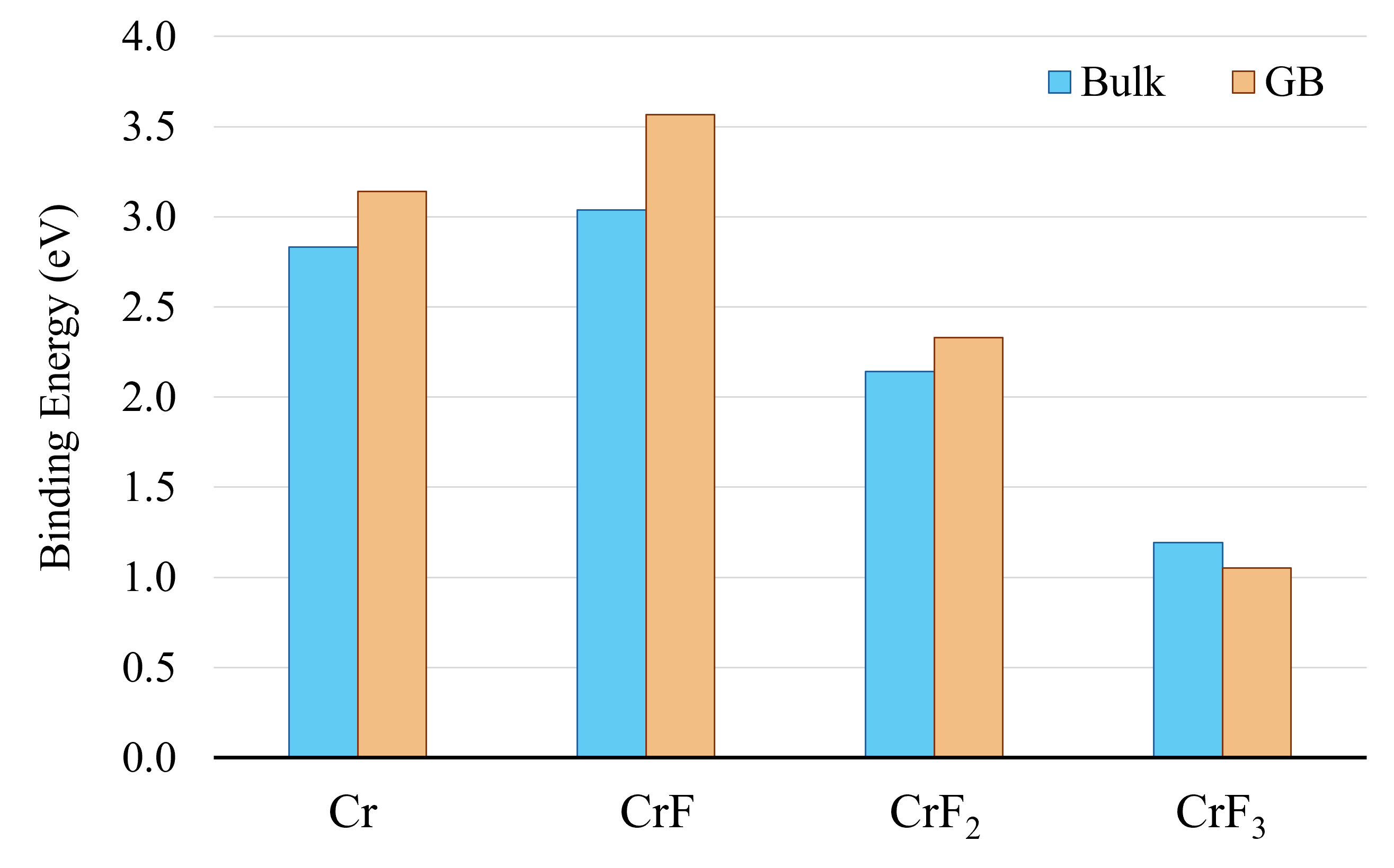}
	\caption{The binding energy of the different CrF$_x$ complexes to the slab surface, at the bulk versus GB.}
	\label{fig:binding_E}
\end{figure}

We first examine the impact of varying fluorine bonding on Cr dissolution behavior, as illustrated in Figure \ref{fig:adatom}. The binding energies of different CrF$_x$ complexes, with $x$ ranging from 0 to 3, adsorbed on both bulk and GB surfaces, are shown in Figure \ref{fig:binding_E}. Previous studies \cite{yin2018first,yin2018theoretical} indicate that bonding with more than three fluorine atoms around Cr is energetically unfavorable, as Cr's binding sites become saturated, inhibiting further aggregation of fluorine. Consequently, configurations with more than three fluorine atoms are not considered in this analysis. The binding energies of various CrFx molecules adsorbed on the surface of a slab, both at the bulk and GB, are shown in Figure \ref{fig:binding_E}. It can be seen that the CrF molecule has the maximum binding energies, though Cr binding energies are not significantly lower than CrF. As the number of fluorine ions further increases (CrF$_x$ $>$ 1), the binding energy decreases significantly. This indicates that the interaction between the CrF$_x$ complex and the surface weakens with higher fluorine coverage. The decrease in binding energy is attributed to the increasing electron affinity and bond polarity introduced by fluorine atoms. Fluorine, being highly electronegative, draws electron density away from Cr, altering the electronic structure and weakening the Cr-surface bond. This effect is particularly pronounced at the GB, where the local atomic arrangement amplifies the destabilization of CrF$_x$ complexes from CrF to CrF$_3$. Particularly, for CrF$_3$, the binding energy at the GB becomes smaller than at the bulk surface. This reduction implies that CrF$_3$ is more susceptible to dissolution at the GB, making it easier for the complex to detach from the GB.

\begin{figure}[!ht]
	\centering
	\includegraphics[width=1.0\textwidth]{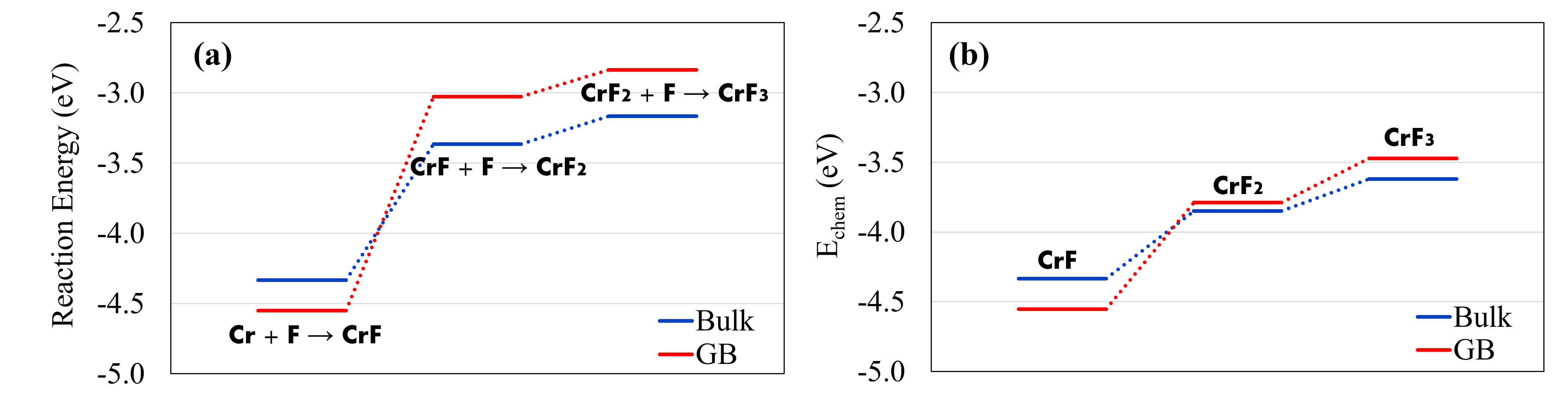}
	\caption{Energy diagrams for sequential fluorination of Cr adatom, comparing bulk and GB regions. (a) Reaction energies for the sequential formation of CrF, CrF$_2$, and CrF$_3$. (b) Chemisorption energy ($\mathrm{E_{chem}}$) for CrF$_x$ complexes.}
	\label{fig:reactions}
\end{figure}

As fluorine atoms progressively weaken Cr's attachment to the surface, the reaction energetics are evaluated to provide insights into the preference for fluorination. The energetic diagrams for the stepwise fluorination of the Cr adatom, depicted in Figure \ref{fig:reactions}a, reveal distinct reaction energies in the bulk and GB regions, including, $\mathrm{Cr+F\rightarrow CrF}$, $\mathrm{CrF+F\rightarrow CrF_2}$, and $\mathrm{CrF_2+F\rightarrow CrF_3}$. Notably, the formation of CrF at the GB has a larger thermodynamic driving force compared to the bulk, suggesting that the GB promotes the initial fluorination of Cr. As the fluorination continues, the enthalpic driving force remains around 3 eV for both CrF$_2$ and CrF$_3$, though slightly weaker at the GB compared to the bulk. Complementing these findings, the chemisorption energies of fluorine, computed based on Eq. (2) and shown in Figure \ref{fig:reactions}b, indicate a higher affinity for fluorine adsorption at the GB during the formation of CrF. However, for CrF$_2$ and CrF$_3$, the fluorine chemisorption energies are more comparable between the GB and bulk, with both regions showing strong fluorine adsorption. These results suggest that while the GB strongly favors the initial attachment of fluorine and CrF formation, subsequent fluorination steps exhibit more similar energetics across GB and bulk regions.

\begin{figure}[!ht]
	\centering
	\includegraphics[width=1.0\textwidth]{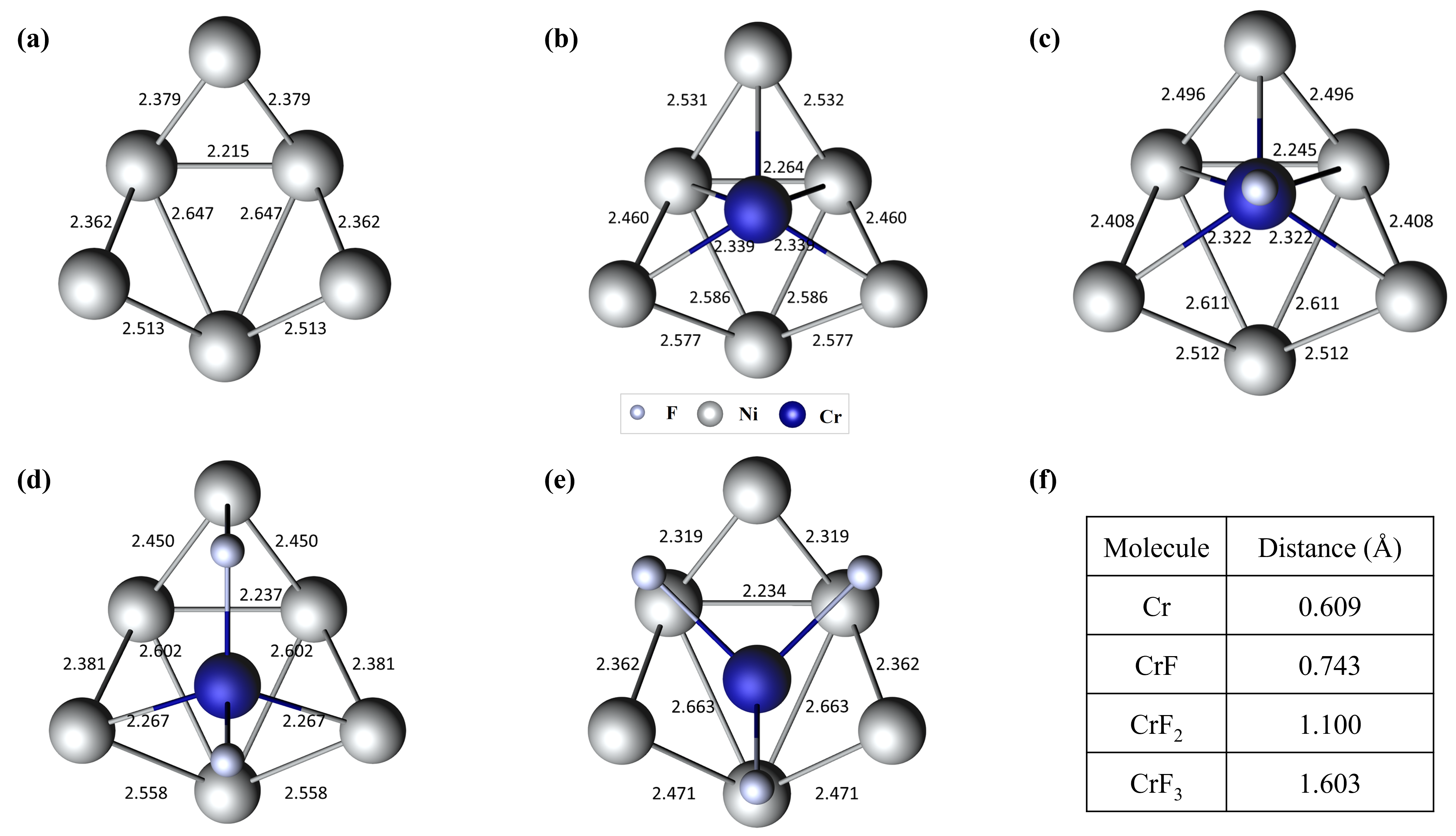}
	\caption{Nearest-neighbor interatomic distances (in $\mathring{A}$) for (a) the pristine $\mathrm{\Sigma}$5(210) Ni GB and bond distances following the adsorption of (b) Cr, (c) CrF, (d) CrF$_2$, and (e) CrF$_3$. Table (f) shows the distance (\AA) between Cr and slab surface, considering the different CrF$_x$ complexes.}
	\label{fig:bonds}
\end{figure}

We next reveal how the bonding environment changes with fluorine adsorption to understand the reaction mechanisms. Figure \ref{fig:bonds} illustrates the nearest-neighbor interatomic distances for the pristine $\mathrm{\Sigma}$5(210) GB and in the presence of CrF$_x$ adsorbed complexes. Cr adsorption at the GB initially leads to an increase in the local Ni–Ni bond lengths, which reflects a disturbance in the metallic bonding due to the introduction of the Cr atom. However, this bond-lengthening effect diminishes as more fluorine atoms bond with Cr. As the Cr evolves from CrF to CrF$_3$, the Cr coordination with the GB decreases, and the distance between Cr and the slab surface increases (Figure \ref{fig:bonds}f); it indicates a progressive weakening of the Cr-GB interaction, which is most pronounced for CrF$_2$ and CrF$_3$. The reduced Cr-Ni coordination and gradual lift-up of Cr suggest that Cr becomes increasingly susceptible to detachment from the GB as fluorine coverage increases. Since CrF$_3$ is least bounded to the GB, Cr is expected to be more prone to dissolution in the form of CrF$_3$.

\begin{figure}[!ht]
	\centering
	\includegraphics[width=0.8\textwidth]{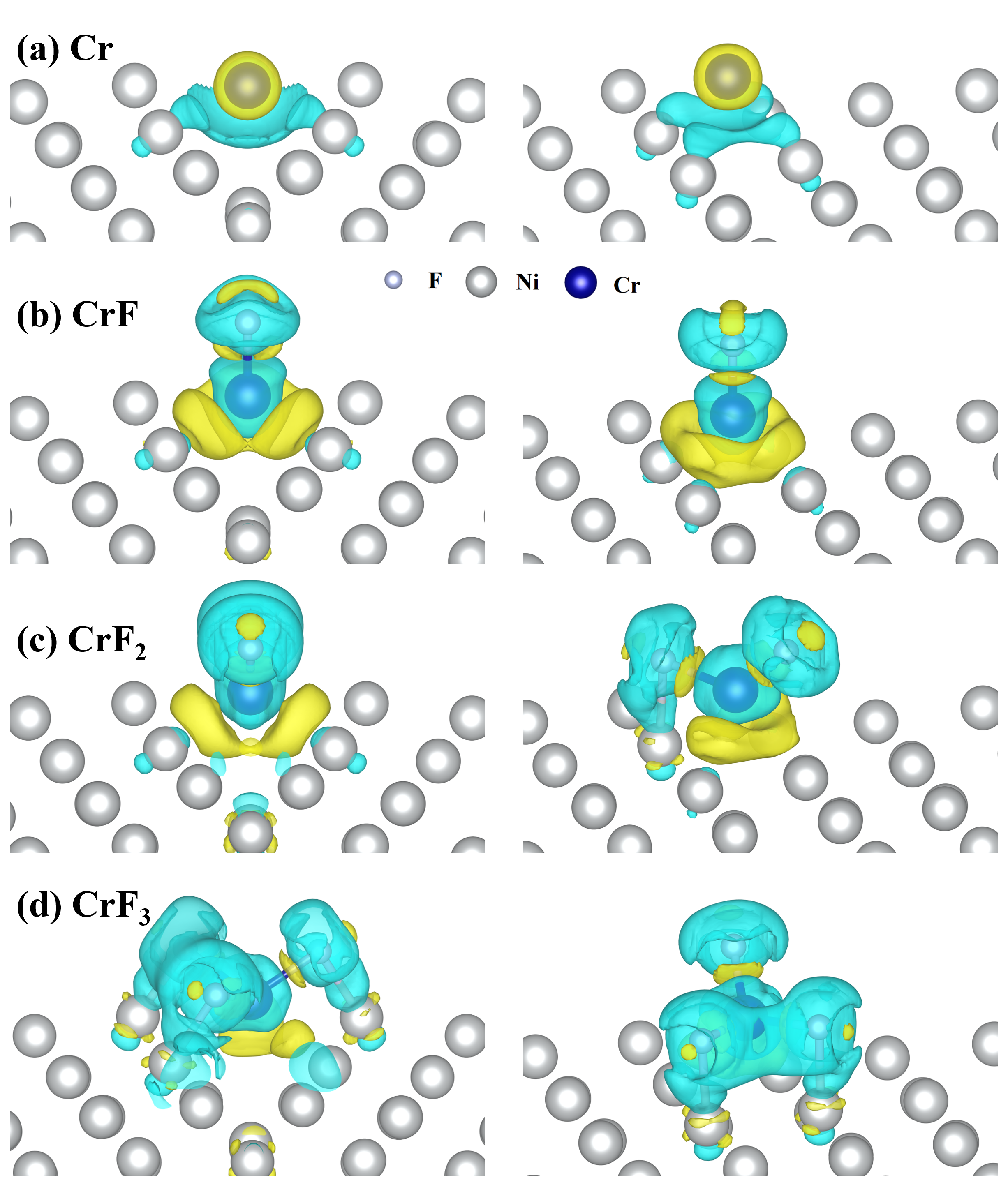}
	\caption{The electronic charge density difference plots for (a) Cr, (b) CrF, CrF$_2$, and CrF$_3$ complexes adsorbed on the Ni(001) surface, comparing adsorption at the GB (left) and bulk region (right). Yellow and blue regions represent charge accumulation and depletion, respectively. The isosurface level is 0.010 e/\AA$^3$.}
	\label{fig:CDD}
\end{figure}

To understand the effects of CrF$_x$ molecule adsorption on the local electron distribution in the adsorbed system, we computed the electronic charge density difference ($\Delta \rho (r)$) as follows: 

\[\Delta \rho (r) =\rho_{\mathrm{CrF_{x}/Ni}} (r)-\rho_{\mathrm{Ni}} (r)-\rho_{\mathrm{CrF_{x}}} (r)\qquad(4)\]
where $\rho_{\mathrm{CrF_{x}/Ni}} (r)$, $\rho_{\mathrm{Ni}} (r)$, and $\rho_{\mathrm{CrF_{x}}} (r)$ are the total electron densities of the adsorbed system, the clean Ni substrate, and the corresponding isolated CrF$_x$ molecule, respectively. All densities are calculated based on their relaxed supercells. Figure \ref{fig:CDD} presents the electronic charge density difference (CDD) plots, where yellow regions represent charge accumulation and cyan regions indicate charge depletion. These plots all reveal significant localized charge redistribution around the adsorption sites regardless of GB or bulk regions, with distinct variations as fluorination levels increase. In the case of the Cr adatom (Figure \ref{fig:CDD}a), charge accumulation near the Cr core indicates the presence of $2p$-orbital electrons close to the Cr atom, with bonding characteristics similar to those observed in Cr-doped Ni metallic systems by Startt et al. \cite{startt2021electronic}. Overall, the Cr adatom is positively charged, with nearby Ni atoms slightly negatively charged (see SM), while the bulk Ni atoms remain largely undisturbed. As F atoms are introduced (Figure \ref{fig:CDD}b-d), the charge distribution undergoes a marked shift. Charge accumulation occurs between the CrF$_x$ and the Ni substrate, while charge depletion becomes concentrated around the adsorbed CrF$_x$ molecules. This indicates a net charge transfer from the CrF$_x$ constituent atoms to the Ni surface. Notably, as the fluorine content increases from CrF to CrF$_2$ to CrF$_3$, the amount of charge accumulation on the Ni substrate decreases. This gradual reduction in charge transfer between CrF$_x$ complexes and the substrate suggests that higher fluorination levels lead to a weaker bond between the CrF$_x$ complex and the substrate. This trend leads to the most ``detached" state for CrF$_3$, where the adsorption strength is significantly weakened compared to CrF and CrF$_2$. 

\begin{figure}[!ht]
	\centering
	\includegraphics[width=1.0\textwidth]{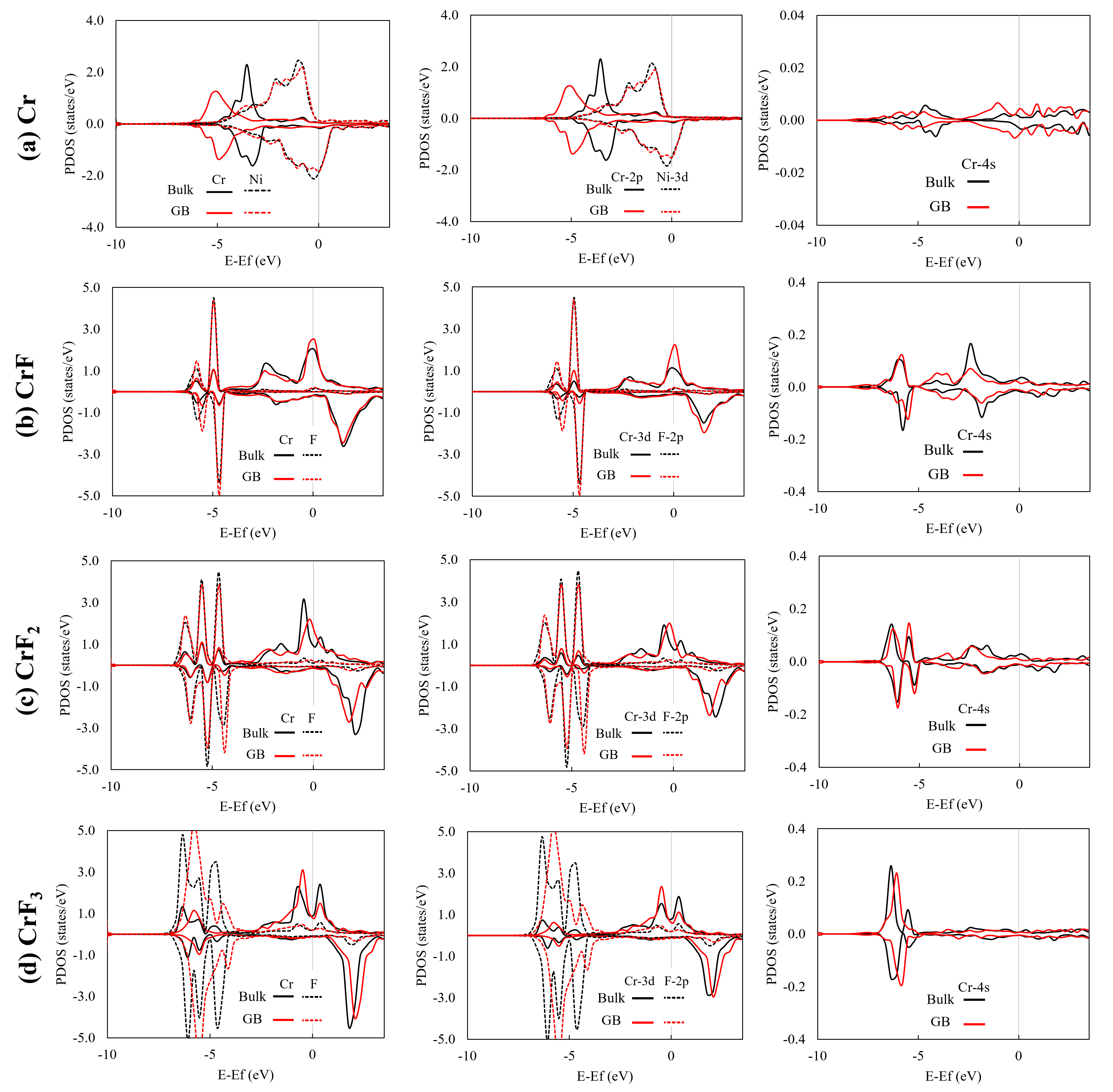}
	\caption{PDOS plots of (a) Cr, (b) CrF, (c) CrF$_2$, and (c) CrF$_3$ complexes adsorbed on the Ni(001) substrate, comparing adsorption at the GB (red) and bulk region (black), where the first column shows the total PDOS plots, the second column shows 3d and 2p orbitals and the third column shows the Cr$-4s$ orbital. The Fermi level is set at 0 eV, indicated by the gray vertical line.}
	\label{fig:PDOS}
\end{figure}

The atomic charges (based on Bader charge analysis \cite{henkelman2006fast,sanville2007improved,tang2009grid,yu2011accurate}) in the adsorbed system reveal that charge transfer between the adsorbate and substrate at the GB is higher than in the bulk (see SM). To understand the difference in the adsorption behavior, the electronic interaction between the Ni substrate and CrF$_x$ molecules is further examined using the projected density of states (PDOS) plots, as shown in Figures \ref{fig:PDOS}. For the Cr adatom (Figure \ref{fig:PDOS}a), the total PDOS is dominated by the Cr$-2p$ orbital contribution. The PDOS peak for Cr adsorption shifts from approximately -3.5 eV relative to the Fermi level (E$_f$) in the bulk to around -4.9 eV at the GB, reflecting a more stabilized electronic state for Cr at the GB. When comparing the PDOS of CrF, CrF$_2$, and CrF$_3$ at the bulk and GB, GB slightly alters the electronic environment around the CrF$_x$ complexes, while its impact on the fundamental bonding interactions is much less pronounced compared to the effects induced by increasing fluorination located at the bulk and the GB. As fluorine atoms are introduced (Figures \ref{fig:PDOS}b-d), the PDOS for both Cr and F reveals notable changes. Specifically, the Cr$-3d$ orbitals begin to exhibit significant overlap and hybridization with the F$-2p$ orbitals, particularly around -5 eV relative to E$_f$. This hybridization indicates a strong interaction between the Cr and F atoms, consistent with previous work on Cr atoms bonded to F on Cr-doped Ni(100) substrates \cite{yin2018theoretical}. In addition, the PDOS for the Cr-4s orbital (Figure \ref{fig:PDOS}b) exhibits a notable peak around -2 eV for CrF, and this peak reduces considerably in the case of  CrF$_2$ (Figure \ref{fig:PDOS}c). Ultimately, it vanishes entirely with the CrF$_3$ complex (Figure \ref{fig:PDOS}d). It indicates that the Cr-$4s$ states account mostly for the bonding between the CrF$_x$ complexes and the substrate, and increasing fluorination weakens the Cr-Ni interaction. Such shift corresponds to the weakening of the Cr-Ni bond observed in the CDD plots (Figure \ref{fig:CDD}). 

\begin{figure}[!ht]
	\centering
	\includegraphics[width=1.0\textwidth]{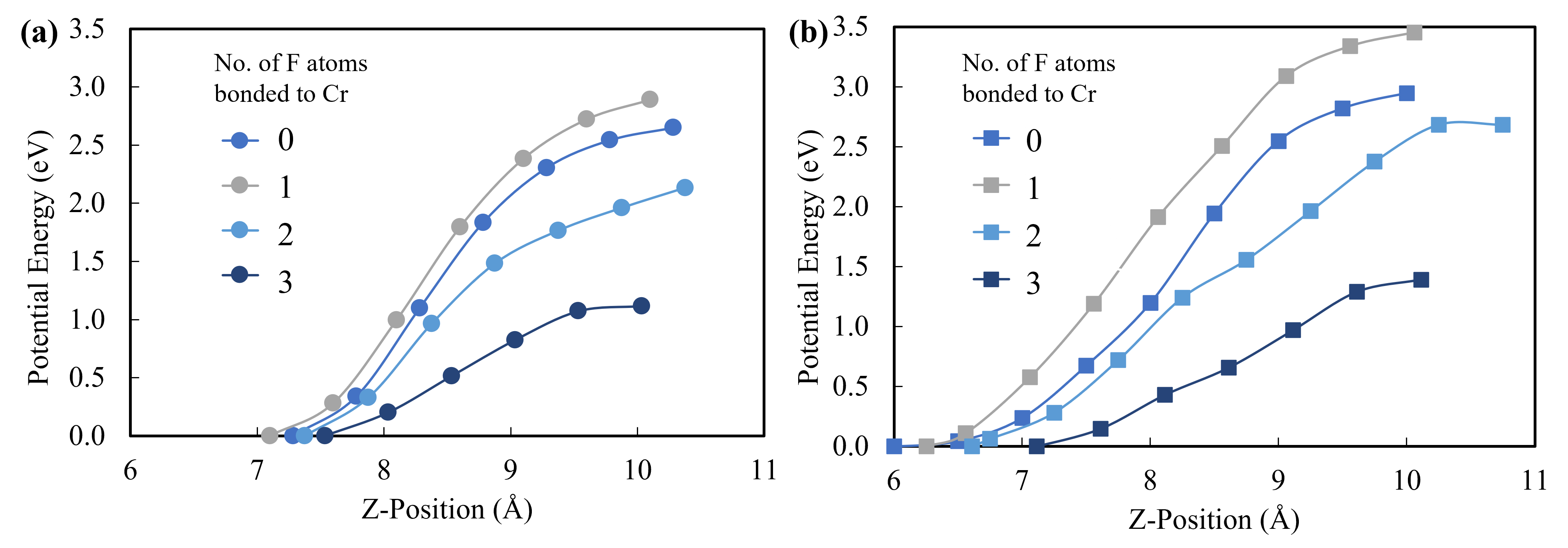}
	\caption{Potential energy profiles for dissolution of a Cr adatom at the bulk (a) and the GB (b), under the impact of adsorbed $x$F atoms. The height is measured with respect to the bottom of the supercell.}
	\label{fig:diss_2}
\end{figure}

To directly quantify the dissolution barrier, the Cr adatom is incrementally displaced from the surface, with up to three fluorine atoms bonded to the dissolving Cr. Figure \ref{fig:diss_2} presents the potential energy profiles as a function of the Cr position during its dissolution, under different fluorine bonding scenarios. Table \ref{tab:table_2} summarizes the dissolution barriers in all the cases. Interestingly, the dissolution barrier increases when a single fluorine atom is bonded to Cr, compared to the fluorine-free case. This initial increase suggests a transient stabilization effect, where the Cr-F bond strengthens the attachment of the Cr atom to the surface. However, when a second fluorine atom is introduced, the barrier decreases substantially. The dissolution barrier is the lowest when three fluorine atoms are bonded to the Cr atom.  This trend is consistent with the binding energy values, shown in Figure \ref{fig:binding_E}. The pronounced reduction in the dissolution barrier for CrF$_3$ highlights its stability as a key corrosion product, where CrF$_3$ forms on the surface before Cr is dissolved, rather than after it enters the salt. It aligns with findings in the literature, where CrF$_3$ is consistently identified in the salt, through both thermodynamic predictions and experimental observations, as the primary corrosion product in fluorine-induced corrosion of NiCr alloys \cite{yin2018first, yin2018theoretical, liu2023temperature}.

\begin{table}[!ht]
    \centering
    \caption{The dissolution barriers for the different CrF$_x$ molecules at bulk versus GB.}
    \label{tab:table_2}
    \begin{tabular}{lll} \cline{1-3} 
    \textrm{Molecule} & Bulk                  & GB                 \\ \hline
    \textrm{Cr}               & 2.650                    & 2.946                 \\ 
    \textrm{CrF}               & 2.890                    & 3.453                \\ 
    \textrm{CrF$_2$}              & 2.133                    & 2.682                 \\ 
    \textrm{CrF$_3$}              & 1.117                    & 1.391                 \\ \hline
    \end{tabular}
\end{table}

When comparing the dissolution barriers for Cr adatoms located at the bulk and at the GB, a similar trend is observed with increasing fluorine adsorption. The dissolution energies at the GB are slightly higher than those observed in the bulk. This might seem counterintuitive, as the GB region is generally expected to have weaker bonding due to its more open and disordered structure. However, the key factor driving this difference lies in the positioning of the Cr atoms. At the GB, the Cr adatom sinks into the boundary region due to the open space, positioning it approximately 1 {\AA} lower relative to the Cr adatom in the bulk, as shown in Figure \ref{fig:adatom}. The Cr adatom dissolution energies at the GB remain comparable to those in the bulk. In the case of CrF$_3$, the dissolution barrier can be reduced to around 1 eV. This observation becomes particularly interesting when we consider two key characteristics: (i) surface diffusion barrier is generally low for adatoms, and (ii) GBs are commonly regarded as high-diffusion pathways for mass transport. It suggests that in the regime of inter-granular corrosion, GB, intercepting the alloy surface, can act as an intermediate ``station" for Cr dissolution, facilitating Cr dissolution directly from GB or from the bulk surface.  

In the presence of a real molten fluoride salt, such as FLiNaK, we expect that the Cr adatom dissolution barrier would be further reduced due to the interactions between the salt and the Cr adatom. The behavior of Cr dissolution in vacuum versus in an ionic or solvent environment is likely comparable to previous studies on metal atom dissolution in the presence of solvents. For example, Taylor \cite{taylor2009transition} simulated the dissociation of a Cu adatom from the Cu(111) surface, and showed that solvent water molecules significantly softened the dissociation process, reducing the energy required for Cu dissociation. Similarly, Ke et al. \cite{ke2020dft} simulated the dissolution of a Ni adatom from the Ni(111) surface and observed a substantial reduction in the dissolution energy barrier with water molecules present compared to vacuum conditions. Hence, it is speculated that the primary mechanism by which fluoride salt reduces the Cr dissolution barrier is through solvation, as the surrounding cation ions can create a strongly solvating environment around the departing Cr adatom, effectively screening the strong Cr-Ni bonds and weakening their interaction. This would stabilize the CrF$_x$ complex as it departs from the surface, minimizing the need for CrF$_x$ to overcome strong metallic Cr-Ni bonds alone as in vacuum conditions. Future work will focus on further exploring the impact of real salt environments on this process.

\section{Conclusion}

This study employs DFT to explore the role of GB in the fluorine-induced initial corrosion of NiCr alloys. By modeling the $\mathrm{\Sigma}$5(210)/(001) symmetrical tilt GB in Ni,  the adsorption and dissolution behaviors of fluorine and Cr atoms were extensively explored. The findings reveal that fluorine atoms preferentially bind to GB sites, and Cr doping further enhances this effect, resulting in stronger adsorption energies at GBs compared to bulk Ni surfaces. By examining the dissolution of Cr atoms in various configurations, both with and without fluorine atoms, we show that the presence of fluorine significantly alters the dissolution dynamics, which is understood from the fundamental analysis of bonding properties. In particular, the formation of the CrF$_3$ complex as a precursor for Cr dissolution notably reduces the energy barrier for Cr dissolution. This implies that as more fluorine bonds with Cr, the Cr atoms become increasingly susceptible to dissolution from the surface. GBs, commonly viewed as structural weak points, can accelerate localized corrosion processes by acting as preferred sites for both fluorine binding and Cr depletion. These findings contribute to a deeper understanding of the early-stage GB-facilitated corrosion mechanisms of Ni-Cr alloys in fluorine-rich environments.

\section*{Acknowledgements}
This work was supported by the National Science Foundation (NSF) under CAREER Award No. 2340019. Any opinions, findings, conclusions, or recommendations expressed in this material are those of the authors and do not necessarily reflect the views of the NSF.

\bibliographystyle{elsarticle} 
\bibliography{references}



\end{document}